# Time-resolved quantitative visualization of complex flow field emanating from an open-ended shock tube by using wavefront measuring camera

Biswajit Medhi,[a,1] Gopalakrishna M. Hegde,[b] Kalidevapura Jagannath Reddy,[c] Debasish Roy,[d],

Ram Mohan Vasu[a]

[a]Department of Instrumentation and Applied Physics,
[b]Centre for Nano Science and Engineering,
[c]Department of Aerospace Engineering,
[d]Computational Mechanics Lab, Department of Civil Engineering,
Indian Institute of Science, Bangalore- 560 012, India
[1]**corresponding author:** biswajitm@iisc.ac.in

**Abstract**

Quantitative visualization of shock induced complex flow field emanating from the open end of a miniaturized hand-driven shock tube (Reddy tube) is presented. During operation, the planar shock wave of Mach number $M_i$=1.33 (±0.6%) is discharged through the low-pressure driven-section, kept open to ambient atmosphere. From the moment of shock discharge, its aftereffects of evolving flow field are recorded for $300 \mu s$ near the exit of the tube by using our newly developed high resolution (16Mpixel) home built wavefront measuring camera setup. The ability of the camera to identify the amplitude and phase of the incident light wave is utilized to measure the flow induced change in phase of the interrogating light beam quantitatively. Information of the evolving flow field with a spatial resolution of 16μm/pixel and time resolution of $50 \mu s$ is recorded in repeated runs. The measured phase information is used in an iterative refraction tomographic scheme to recover the 3D-density distribution of the flow field, which reveals the internal features of the domain. Computational fluid dynamic (CFD) simulation is carried out for the same experimental conditions and it is found that recovered experimental density distribution shows good agreement with the results obtained through CFD simulation.

## 1. Introduction

The invention of shock tube (ST) in 1899 by Paul Vielle [1] opened a new era that taught the way of harnessing the tremendous strength of shock wave and reproducing it in the safest and repeatable manner. ST and its extension, shock tube driven shock-tunnels are extensively used in generating high speed, high enthalpy gas flows in supersonic and hypersonic regimes [2] for aerodynamic investigations. The ability to change the local property of gas instantaneously by introducing shock to it in an effortless yet reliable way is the key feature of attraction towards the different applications of ST [3-5]. Among the various applications of ST, either the generated primary shock or the shock discharged from its open end finds attention [3-5]. Besides its fundamental importance in the scientific community, shock wave diffraction phenomenon finds technical applications even though it is in the verse of understanding. When shock wave is discharged from a ST to the ambience it creates an unsteady complex flow field that includes expansion and diffraction of shock followed by formation of shear layer and vortices, the formation of embed shock, trailing jet, Kelvin-Helmholtz flow instabilities etc. Other than these mentioned aftereffects shock tubes have also been employed to study interactions among shocks, shock-body, shock-vortex structures [6-10] and generated acoustic noise [11] due to these interactions, leapfrogging phenomenon of vortex rings [12], effect on out coming flow due to exit shape of nozzles [9, 10, 13-15] and many more.

Investigations of flow fields using optical techniques are always preferable since they are non-intrusive and carry useful information [6-39]. Their application ranges from providing high contrast qualitative images to the quantitative information about the important fluid properties such as density, temperature, pressure, velocity etc. Many optical techniques such as schlieren, shadowgraph, holographic interferometry, particle image velocimetry (PIV), background oriented schlieren (BOS) [17- 20], shadow-casting [21, 22], laser induced fluorescence (LIF) [17, 18] are extensively used in flow visualization in supersonic and hypersonic regimes in shock tunnels and wind tunnels. Among these, schlieren and shadowgraph schemes provide 2D projected qualitative images of the 3D domain, represented in intensity scale. Here, images are formed by the interrogating beam whose deflection is proportional to first- and second order derivative of the refractive index of the medium through which ray passes. Studies related to shock discharge from open-end of ST have been interrogated qualitatively by both the schemes; starting with one of the first attempts made by Edler and de Hass [23] followed by several other researchers [6-9, 11-14, 23-25]. For quantitative flow visualization, holographic interferometry is a promising technique; in which deformed fringe pattern encodes the information of the wavefront coming out of flow that further processed to recover density distribution. Recently holography has been employed to study shock- shock, shock-vortex interactions and is provided in the references [10, 11, 25- 27] in detail. PIV technique, which was originally developed targeting incompressible flows of low Mach number, is also implemented in flow evolution studies close to the exit of a ST [28-31]. By tracing the seeding particles present in the flow, local velocity distributions within the flow field are obtained. Another promising technique based on BOS, in which the refractive gradient induced distortion of the background during the presence of flow is used as a measure for qualitative/quantitative evaluation of the flow field and its application in ST studies can be found in references [31, 32]. However, most of the above-mentioned techniques require either an expensive camera, pulsed laser or complicated optical set-up to extract quantitative information about the flow fields. Hence, there is need to develop a simple, inexpensive and rugged experimental technique for quantitative flow visualization of high-speed flows. Present work is an attempt that addresses the above-discussed issues in optical flow diagnostics for quantitative recovery of flow fields.

In this paper, quantitative visualization of shock induced flow field discharged from the open end of a shock tube is presented, in which the discharged flow field is investigated through our recently developed wavefront measuring camera [36, 37]. For the set of experiments, a miniaturized, table-top shock tube, also known as Reddy tube [34, 35] is used, which shares the similar working principle as by the Stalker tube [33]. During experiments, mentioned ST is driven manually to increase the pressure of the driver side by using a piston arrangement until it breaks open the primary diaphragm creating planer shock wave, which undergoes diffraction on release through the exit of the ST to the ambient. The diffracted shock at different instant of time is investigated quantitatively by our quantitative wavefront measuring camera set-up [36, 37] by capturing conjugative images of the flow field with a spatial resolution of 16μm/pixel and temporal resolution of $50\mu s$ for the first $300\mu s$ of flow evolution. Captured images are processed to recover density distribution of the flow field, which are further analyzed and compared with the one obtained through computational fluid dynamic (CFD) simulation. Flow features such as shock front, vortex ring, shear layer, embedded shock are prominent in the recovered results. The details of the experimental facility, the in house camera, experimental procedure, and recovery of results through experiment and CFD simulations are discussed in the following sections.

## 2. Experimental setup and procedure

### 2.1 Experimental setup

Experiments are conducted using the single diaphragm hand-operated Reddy tube (see Fig. 1) [34, 35]. It has 31*mm* inner diameter for both driver and driven sections, of lengths 400*mm* and 600*mm* respectively. Four pressure transducers are attached to the ST, among which one is placed in the driver section for measuring the diaphragm rupture pressure and the rest of the three are in driven section respectively. Attached three pressure sensors with sensitivity of 6.4mv/kPa each, are placed at *20, 170, 320mm* away from the open end of the tube respectively, which is shown schematically in Fig. 1(a). A primary diaphragm made of tracing paper (*90– 95 g.m$^{-2}$*) separates the driven section from the driver side, where another end of the driven side is left open to the ambient conditions (*$P_1$=680mm* of Hg, *$T_1$=300K*). During operation, the attached piston of the driver side is pushed forward increasing the driver pressure until it ruptures open the primary diaphragm. By setting the thickness of the diaphragm, rupture pressure of the primary diaphragm $P_{TH}$, which is $P_4$, and the effective driver length, $L_{eff}$ (see Fig. 2), the ST can be tuned to attain different incident shock Mach number (*$M_i$*). In the present case, single layer primary diaphragm helps to build $P_4$~4.1bar, $L_{eff}$ ~9.5cm and *$M_i$*=1.33. An insert is also placed in-between the piston handle and driver section to maintain the $L_{eff}$ fixed, increasing the probability of repeated runs. Diaphragm bursting generates primary shock wave that moves forward towards the downstream in the driven section and exits through the open end of the ST to the ambience. Discharged shock and its aftereffects close to the exit of the ST are intercepted by the optical imaging system as shown in the schematic of Fig. 2. It mainly consists of an interrogating (plane wave) beam derived from a quasi-monochromatic light source, *S* and a wavefront measuring camera. Quantitative visualization of the flow field is accomplished by using the wavefront measuring camera [36, 37] that can measure the phase of the interrogating beam, although its current hardware doesn't support high frame rate recording matching the need to image the dynamic flow field at the exit of the ST. Therefore, images of the flow field is captured in a series of identical experiments (one image at a time) by triggering the camera unit at different time instances covering the first 300μs time of flow evolution after the shock discharge from the ST.

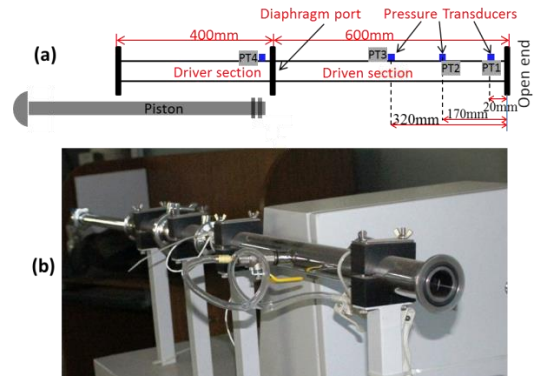

Fig.1 Reddy tube (a) Schematic, (b) Actual setup

Reproducibility of the experiments is must to cover the dynamics of evolutionary flow while capturing different instances at different experimental runs such that the stitched images from different experiments resemble an evolving flow field corresponding to a single experiment if it were captured by a single high-speed camera. In the present experiments by using our camera set-up this is achieved by: 1) maintaining the rupture pressure $P_4$ fixed, 2) triggering the camera unit at different delayed sequences to capture the flow field at different time instances. Fixed $P_4$ ensures a stable incident shock Mach number ($M_i$) of the ST, which is maintained at 1.33, a figure repeatable to within $\pm 0.6\%$, calculated from the pressure signals acquired from the sensors of the driven section. One of these sensors is also used for triggering the camera unit during experiments. With the specified $M_i$, for a given time instant, say $100\,\mu s$, several experiments are conducted and the captured images are analyzed frame by frame to get the position $Z_s$, of the diffracted shock along the tube axis (see Fig. 3). Runs that satisfy both $M_i$ and $Z_s$ within the specified limit are used for further studies; where $Z_s = \pm 1$ pixel of the digital image is used as a selection criterion of $Z_s$.

**2.2 Optical arrangement and the quantitative wavefront measuring camera**

A 4-f optical setup (see Fig.2), which is common for schlieren imaging is used to interrogate the shock induced flow field discharge from the ST. Interrogating beam derived from a quasi-monochromatic light source $S$ ($\lambda = 530\,nm$) is collimated by a concave mirror $M_1$ to illuminate the flow field, which further collected and directed to the camera with the help of concave mirror $M_2$ and plane mirror $M_3$ respectively. All the mirrors are front coated with surface finishing of $\lambda/4$. In the current study, we have used our recently developed wavefront measuring camera, which is placed after the second mirror of the optical setup without any knife-edge or color filter, focusing on a plane very close to the ST surface such that it sees the light wave exiting the flow field. Flow induced phase deformation of the interrogating beam is detected quantitatively by the wavefront camera which is beyond the reach for a normal camera (intensity detector). Images of the flow field are captured at different time instances in different experiments (identical runs), which are processed to recover quantitative density distribution of the expanding flow fields in time. The details of data processing to ascertain density from the captured images is discussed in the next section. A brief description of the quantitative wavefront measuring camera is given below.

**2.3 Wavefront Camera:**

The quantitative flow field measurements described in this manuscript is based on the home built *quantitative wavefront measuring camera* [36, 37]. The camera is built around a modified CMOS image sensor embedded with a semi-transparent mask positioned facing the sensor in close proximity, which provides the ability to detect the direction of the incident light wavefront. Under geometrical optics, the casted shadow of the mask on the sensor plane is the eikonal approximation of the captured light wavefront through the mask pattern and thus any variation of the captured wavefront would displace the casted pattern from its original position when illuminated by a plane wave. The observed phenomenon is similar to the shadow-casting process, which has been reported before [21, 22]. By processing the captured images of the flow field directional slopes of the wavefront is retrieved quantitatively; which further used for tomographic density recovery of the flow field by following the procedures as explained in the sections below.

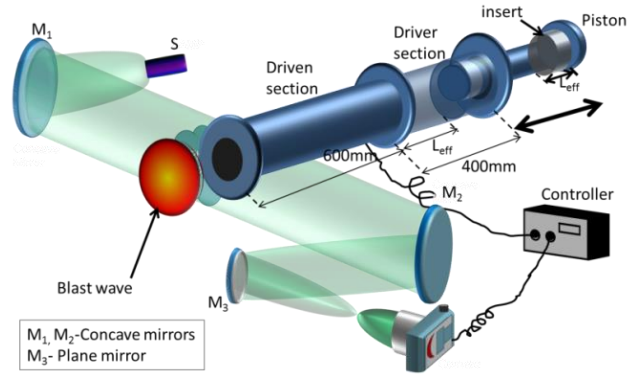

Fig. 2 Schematic of the optical setup and the camera used.

**3. Data processing and tomographic reconstruction of density**

Flow induced change in refractive gradients distort the interrogating beam and the variations of the exit beam from the flow field is recorded as the local slopes of the wavefront by the wavefront measuring camera. For a smooth recovery of the wavefront $\Phi(x, y)$ from its local slopes $[\theta_x, \theta_y]$, the error $\varepsilon^2$ of Eq. (1) is minimized [21, 22, 36, and 37]

$$\varepsilon^2 = \sum_{i=0}^{K-2}\sum_{j=0}^{L-1}(\Phi_{i+1,j} - \Phi_{i,j} - \theta_{i,j}^x)^2 + \sum_{i=0}^{K-1}\sum_{j=0}^{L-2}(\Phi_{i,j+1} - \Phi_{i,j} - \theta_{i,j}^y)^2 \qquad (1)$$

Here $K$ and $L$ are the dimensions of $\Phi(x, y)$ in discretized computational domain. The process of recovering field parameter (density in this case) from projections belong to the broad class of inverse problems in which tomographic reconstruction algorithms centered around series expansion and transform based schemes [39] are used. Here, the unwrapped phase $\Phi(x, y)$ obtained through Eq. (1) is used for recovering refractive index $\mathbf{n(r)}$ of the medium by using the iterative refraction tomography, in which the following inverse problem is solved [21]:

Find $\mathbf{n}(\mathbf{r})$, such that $\quad\quad \Phi_i = \frac{2\pi}{\lambda} \int_0^L n(i) dl_i \quad\quad$ i=1…total no. of rays $\quad\quad$ (2)

Integral of Eq. (2) is along the Fermat's path connecting source and detector by obeying the eikonal equation Eq. (3)

$$\frac{d}{ds}\left(\mathbf{n}\frac{d\mathbf{r}}{ds}\right) = \nabla \mathbf{n} \quad\quad (3)$$

Recovered refractive index distribution, $\mathbf{n}(\mathbf{r})$ in the flow field is converted to density by using Gladstone-dale equation Eq. (4) as mentioned below

$$\rho = \frac{(n-1)}{K_\lambda} \quad\quad (4)$$

here $\rho, n$ and $K_\lambda$ are recovered density, refractive index and Glad-stone dale constant respectively

The used ST has circular symmetry at the exit and therefore the axis-symmetric assumption is made while performing tomographic reconstruction from the recorded projections. For a given instant of time, a captured image that satisfy both the qualifying criteria as mentioned in section 2.1 is used to process for recovering slopes of the wavefront followed by wavefront recovery and finally the density by phase tomography.

## 4. Numerical computational fluid dynamic (CFD) simulations

Numerical simulations are performed that help in understanding the flow evolution phenomenon and also used for assessment of the recovered experimental results obtained through the wavefront measuring camera. Shock wave discharge from the open end of a ST to the ambience is numerically simulated by using the commercial CFD software ANSYS FLUENT 13$^{(R)}$ [40]. Navier-Stokes equations are solved for a single component viscous gas flow by using finite volume method in Cartesian coordinate system. Transient model is used for simulating the time-dependent nature of the flow evolution process, in which convective fluxes are calculated at the interface of control volume by Roe-FDS scheme [40]. The primitive variables are interpolated at the interface with the help of second order upwind approach [40]. Used numerical method is based on Implicit formulation of the problem where transient part is handled through second-order implicit scheme [40].

Experimentally measured incident shock Mach number ($M_i = 1.33$) has been utilized to set the initial conditions for conducting the simulation. Driver and driven gases are set to air, and the initial $P_4$ and $T_4$ values are set to *4.1bar* and *310K* respectively. Used test gas is assumed to be calorically perfect with specific heat constant $\gamma = 1.4$. For the viscous simulation, the viscosity is calculated from Sutherland's law. The used shock tube has circular symmetry and therefore, 2D axis-symmetric model is used while evaluating the CFD simulations, resembling the experiments.

## 5. Results and discussion
The schematic of the experimental setup is shown in Fig.2. For this quantitative flow visualization, the wavefront measuring camera [36, 37] is used to capture the flow field images with spatial resolution of 16µm/pixel and exposure time of $0.8\mu s$. The camera unit is triggered by the signal derived from one of the pressure transducers of the driven section with the help of a control unit (see Fig.2), which controls the timing of triggering signals used in different repeated experiments. Repeatability of the experiments is assured by maintaining the incident shock Mach number fixed to 1.33 and the position ($Z_s$) of the diffracted shock front along the tube axis which is discussed in detail in section 2.1. Captured images of the flow field at different time instances are processed to recover density gradients, which are used later to recover density distribution in the flow field. The measured density gradients of the flow field shown in Fig.3, resemble the schlieren images of the same while using vertical (Row 1) and horizontal (Row 2) knife edge respectively. Flow induced wavefront deformation of the interrogating beam are shown in the last row of Fig. 3, obtained after following the procedure described in section 3.

In experiments, due to sudden rupture of primary diaphragm compression waves are generated and they coalesce to form the planar shock wave which propagates towards the downstream to the exit of the tube, whereas generated expansion waves move towards the end of the driver section. The planar shock front emanates first followed by the test gas, contact surface and reflected expansion wave from driver section respectively. The exit shock at the moment of its appearance to the ambient is shown in the first column of Fig. 3 and is labeled as $\tau\ \mu s$. The shock wave diffracts to the ambient which was planar before, loses its strength with the increase in angle to the shock tube axis and can be seen from the obtained density gradients map of Fig. 3. Velocity difference between the ambient and the flow exiting from the tube initiates the development of shear layer and that is seen from density gradients of Fig. 3, especially from the time instant $(\tau + 150)\mu s$ onwards. The size of shear layer increases while moving along the flow direction and finally starts rolling into counter rotating vortices. Circular symmetry of the tube-exit imposes the symmetry on the generated vortex ring which is evident from the reconstructed density profile of Fig. 4. The diffracted shock front travels at much faster speed than the created vortices in the flow field. In the presented results, generation of vortices can be seen first at $(\tau + 50)\mu s$, which accelerates along the flow direction up to $(\tau + 150)\mu s$, then decelerate $((\tau + 150)\mu s$ onwards) that is evident from Fig. 3. From the images it can also be seen that the size of the vortices which is apparent at the upper and lower corner of the tube exit increases in size along its travel path. The core of the vortex ring is at the center of dark and bright regions of horizontal gradient. Embedded shock (ES) evident from the vertical gradients of Fig. 3 is sitting along with the vortex rings and that is responsible for slowing down the high speed gas coming out from the tube to the ambient.

For the quantitative recovery of density using tomographic schemes, data from time instants of $(\tau + 50)\mu s$ and $(\tau + 100)\mu s$ of Fig.3 are considered. Calculated 3D-density profile of the flow field is shown in Fig. 4, in which the 3D contour of shock front and the vortex ring are distinguishable resembling the shape of diffracted shock front with a mushroom like structure. The center slice of the reconstructed density profile in Fig.4 is shown in comparison with the CFD results for the same experiment. The comparison is shown separately in Fig.5 which clearly shows the agreement of experimentally recovered density profile to the one obtained from CFD simulations. However, the shear layer is not distinguishable for the experimentally recovered data but is visible from the CFD simulation. The suppression of such information could be due to smoothing process during phase recovery while solving the minimization given by Eq. 1. Moreover, the time averaged recorded data of finite exposure of $0.8\mu s$ smoothens the crisp information present in the flow field.

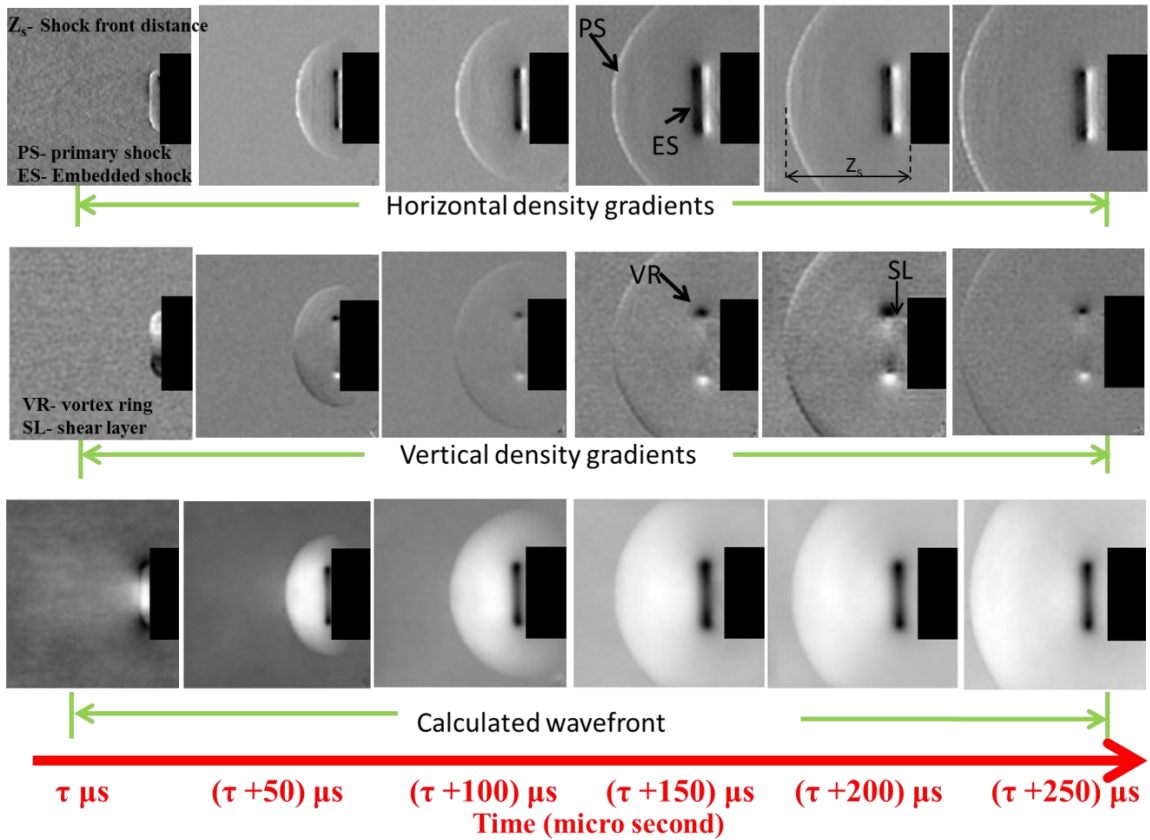

Fig. 3: Flow evolution from the open end of Reddy tube; Row1- Horizontal gradient, Row2- Vertical gradient, Row3- recovered phase

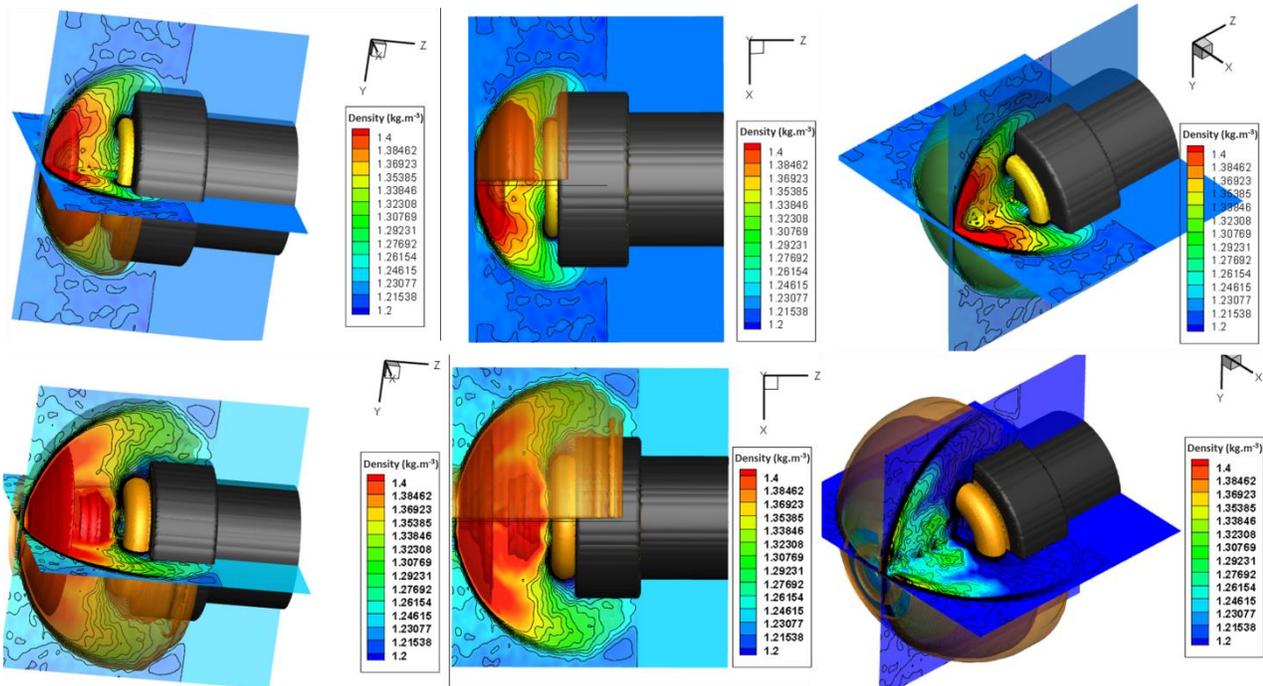

Fig. 4 Reconstructed 3D-density profile of the flow field rendered with different orientations. 3D-density map by tomographic inversion at $t = (\tau + 50)\mu s$ (Row1) and $t = (\tau + 100)\mu s$ (Row2)

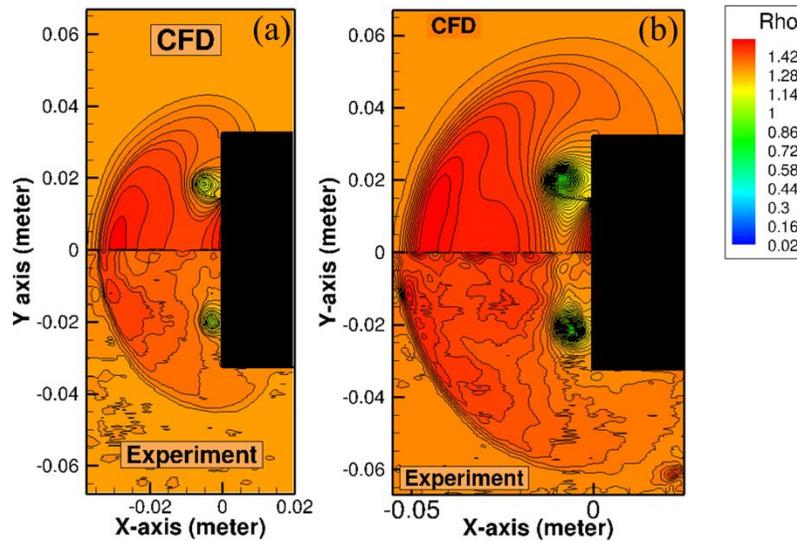

Fig.5 Comparison between experiment and CFD (a) at $(\tau + 50)\mu s$ and (b) $(\tau + 100)\mu s$ respectively. Upper portion of Figs (a) and (b) represent the CFD simulated results and the lower half corresponds to experiment

Internal features of the reconstructed 3D-domain is extracted by representing the 3D profile in to 2D cross sections at different Z/D ratios, where Z and D corresponds to the travel distance of the shock front from the exit of the ST and the inner diameter of the tube respectively. The slices are taken for $Z/D$ of 0.025, 0.05, 0.075 and 0.01 for the frame at $(\tau + 50)\mu s$; and 0.05, 0.01, 0.15 and 0.2 for the frame at $(\tau + 100)\mu s$ respectively. The obtained density slices at the mentioned $Z/D$ ratios for time instances $(\tau + 50)\mu s$ and $(\tau + 100)\mu s$ are shown in Figs.6 and 7 respectively. The size of the mushroom cloud expands with time and thus the obtained circular zones of Figs. 6 and 7. It is seen form the reconstructed density slices that the shock-wave induced regions within the mushroom cloud are of higher density regions than the surroundings; whereas the region inside the vortex ring is at lower value.

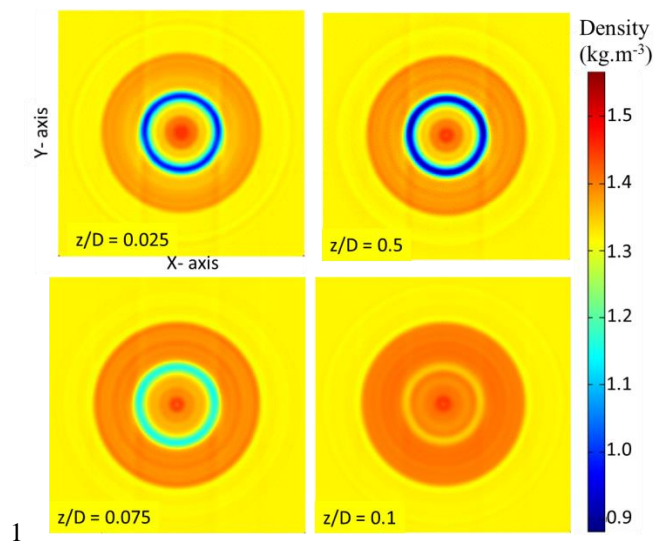

Fig:6. Cross-sectional recovered density at $(\tau + 50)\mu s$ at different Z/D ratios (0.025, 0.05, 0.075 and 0.01)

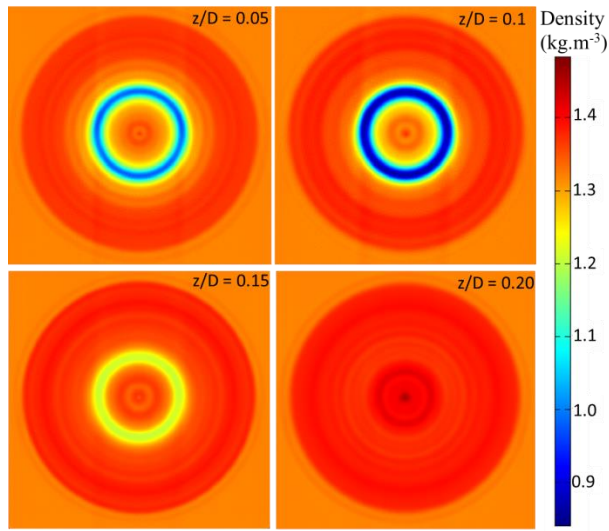

Fig:7 Cross-sectional recovered density at $(\tau + 100)\mu s$ at different Z/D ratios (0.05, 0.01, 0.15 and 0.2)

## 6. Conclusions

In conclusion, quantitative visualization of flow field by using a home built wavefront measuring camera setup is presented. In the present case, time resolved flow field emanating from a ST is investigated for quantitative density measurements. The evolving flow field is interrogated by the wavefront measuring camera with a spatial resolution of 16μm/pixel and 50μs temporal resolution between two consecutive frames. Flow induced change in refraction that leads to wavefront distortion of the interrogating beam coming out of the flow field is detected by the stated camera unit quantitatively, which later used as an input to the tomographic reconstruction of the flow field. Estimated 3D-density distribution of the flow field reveals the internal distribution of the flow field, containing diffracted shock, shear layer, vortex ring etc. It is also found that the forward facing and backward facing shocks are of different density profiles. Vortex ring generated at the exit of the tube accelerates first (up to 150μs in this case) then slows down; moreover, the size of the vortex also increases with time unless it dissipates. Experimentally measured density profiles show a good agreement with the simulation results of the flow field obtained through CFD. The developed technique of quantitative density measurement with the help of a wavefront measuring camera shows promising results and could be extended to other flow dynamics studies.

## Acknowledgements

We acknowledge the financial support from DRDO, India.